# A Multimodal Ecological Civilization Pattern Recommendation Method Based on Large Language Models and Knowledge Graph

Zhihang Yu[1,3], Shu Wang[2*], Yunqiang Zhu[2], Zhiqiang Zou[1,3, *]

[1]College of Computer, Nanjing University of Posts and Telecommunications, Nanjing 210023, China.

[2]State Key Laboratory of Resources and Environmental Information System, Institute of Geographic Sciences and Natural Resources Research, Chinese Academy of Sciences, Beijing, China.

[3]Jiangsu Key Laboratory of Big Data Security & Intelligent Processing, Nanjing University of Posts and Telecommunications, Nanjing 210023, China.

* Corresponding author:

Shu Wang ( wangshu@igsnrr.ac.cn )

Address: 11A, Datun Road, Chaoyang District, Beijing, 100101, China;

Zhiqiang Zou(zouzq@njupt.edu.cn)

Address: 11A, Wenyuan Road, Qixia District, Nanjing, 210023, China;

*Author emails:*

Zhihang Yu: 1222045518@njupt.edu.cn ;

Shu Wang: wangshu@igsnrr.ac.cn ;

Yunqiang Zhu: zhuyq@lreis.ac.cn ;

Zhiqiang Zou: zouzq@njupt.edu.cn ;

# Abstract：

The Ecological Civilization Pattern Recommendation System (ECPRS) aims to recommend suitable ecological civilization patterns for target regions, promoting sustainable development and reducing regional disparities. However, the current representative recommendation methods are not suitable for recommending ecological civilization patterns in a geographical context. There are two reasons for this. Firstly, regions have spatial heterogeneity, and the (ECPRS)needs to consider factors like climate, topography, vegetation, etc., to recommend civilization patterns adapted to specific ecological environments, ensuring the feasibility and practicality of the recommendations. Secondly, the abstract features of the ecological civilization patterns in the real world have not been fully utilized., resulting in poor richness in their embedding representations and consequently, lower performance of the recommendation system. Considering these limitations, we propose the ECPR-MML method. Initially, based on the novel method UGPIG, we construct a knowledge graph to extract regional representations incorporating spatial heterogeneity features. Following that, inspired by the significant progress made by Large Language Models (LLMs) in the field of Natural Language Processing (NLP), we employ Large LLMs to generate multimodal features for ecological civilization patterns in the form of text and images. We extract and integrate these multimodal features to obtain semantically rich representations of ecological civilization. Through extensive experiments, we validate the performance of our ECPR-MML model. Our results show that F1@5 is 2.11% higher compared to state-of-the-art models, 2.02% higher than NGCF, and 1.16% higher than UGPIG. Furthermore, multimodal data can indeed enhance recommendation performance. However, the data generated by LLM is not as effective as real data to a certain extent."

# 1. Introduction

The ecological civilization pattern (namely sustainable development pattern) refers to a pattern that emphasizes the harmonious coexistence and sustainable development of humans and nature. It emphasizes that in the process of development in various fields such as the economy, society, and culture, humans should adhere to the principles of ecological priority and green development. This involves protecting and restoring the stability and health of ecosystems to achieve a harmonious coexistence between humans and nature. In other words, efficiently and accurately recommending the ecological civilization pattern is one of the pathways to achieving Sustainable Development Goals (SDGs). However, recommending the ecological civilization pattern is different from recommending commodities. It possesses geographical spatial heterogeneity and a complex, diverse nature, presenting significant challenges in constructing ECPRS. Therefore, the algorithm for recommending ecological civilization patterns is crucial for application in the field of sustainable development and represents a core bottleneck issue for the intelligence of sustainable development.

Currently, recommendation algorithms that can be used for ECPRS can be categorized into two groups, the first being classical commodity recommendation algorithms and the second being algorithms that take into account geospatial heterogeneity. The main idea of the first group of methods is to utilize the features of the commodity side, such as in the KGAT(Xiang Wang, He, Cao, Liu, & Chua, 2019)approach where a movie, as an item, utilizes information about the movie's director, producer, main cast, etc., to construct a knowledge graph and extract features to enhance the vector embedding representation of the item, and to recommend for the user the movie that the user may like. In the ecological civilization pattern recommendation scenario, the user is the region and the item is the ecological civilization pattern, the ecological civilization pattern doesn't have these real-world features, and by ignoring the large number of features such as soil, water and air on the region side, i.e., ignoring the geo-spatial heterogeneity features. Therefore, the first group of methods is not suitable for ECPRS. To address this issue, the second group of methods has been tailored for ECPRS, detailed information about such methods, please refer to section 2.1 of the related work. The main idea of these methods is similar to that of the first group, which is to extract the features of the regions so as to take into account geographical spatial heterogeneity. Although the performance of ECPRS is improved compared with the first category of methods, these methods have a common disadvantage that they ignore the abstract features of the ECPRS's item, i.e., the ecological civilization pattern. As shown in Figure 1, they use the knowledge graph to extract the features of regions, while the ecological civilization model is embedded only using its corresponding number, and not considering the features of the ecological civilization pattern may lead to the difficulty of the recommender system to provide personalized recommendations. Regions' preferences may be associated with attributes, content, or features of a particular item, and if these features are not considered, recommendations will not be customized based on regions' hidden preferences.

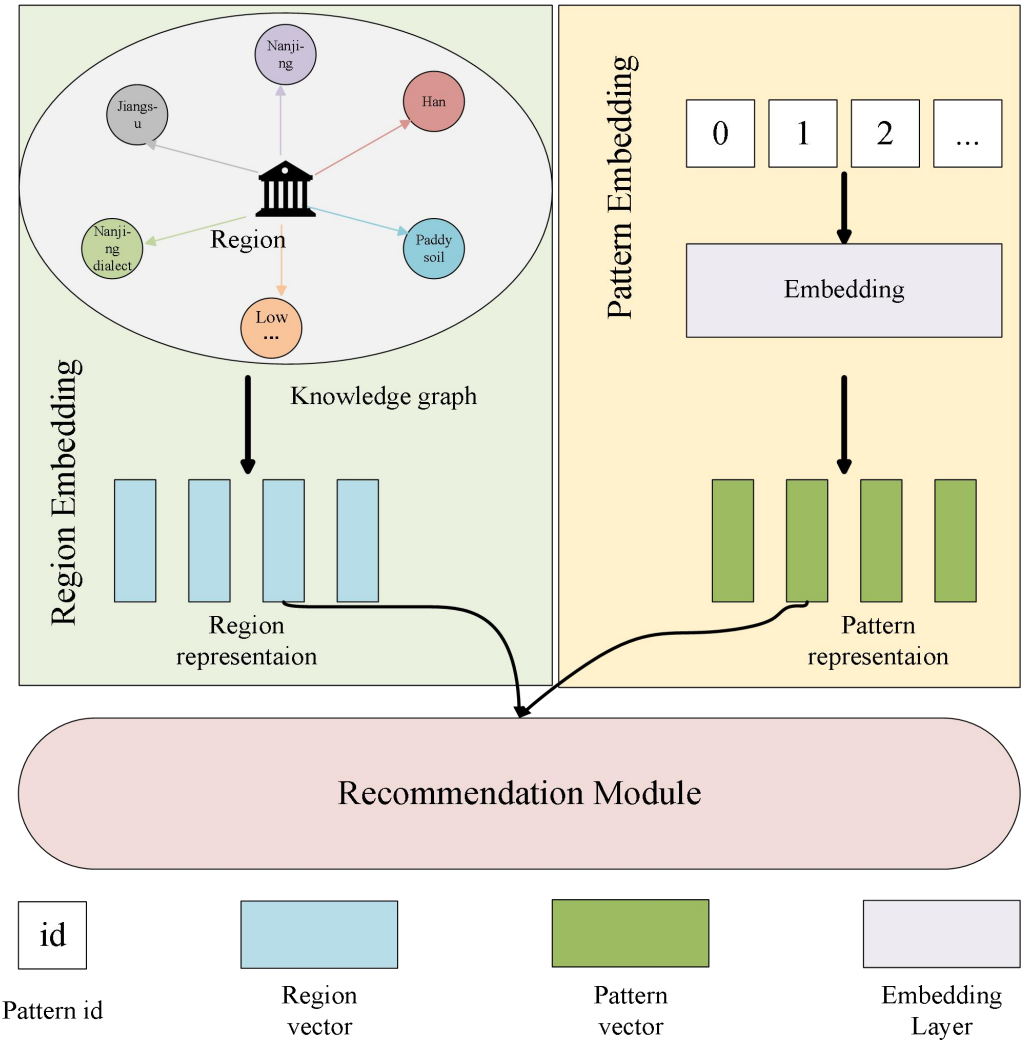

Figure 1 Embedding process of region and pattern

Considering this limitation, the aim of this study is to increase the semantic information contained in the ecological civilization pattern representation so that it represents the characteristics of the ecological civilization pattern as comprehensively as possible. Currently, the reason for the lack of rich semantic information in ecological civilization pattern vectors is that the patterns themselves do not possess real-world features like color or brand, which could be used to enrich their semantic information. Fortunately, large models provide a new avenue. We can leverage the outstanding semantic understanding capabilities of large models to generate multimodal data for ecological civilization patterns. From this data, we can extract and integrate features of ecological civilization patterns to obtain representations that encompass abundant semantic information. Therefore, combining with UGPIG, we propose the method A Multimodal Machine Learning Method for Ecological Civilization Pattern Recommendation (ECPRMML). The main idea of this method is to utilize the extracted semantic information from the knowledge graph to represent embeddings for the regions, extract features from the ecological civilization pattern multimodal data generated by Large Language Models (LLMs) and fuse them to form the representation of the ecological civilization patterns. With these two representations, we can calculate and obtain the final recommendation results. The contributions of this paper can be summarized as follows:

- To the best of our knowledge, this is the first endeavor to simultaneously integrate knowledge graphs, large language models, and multimodal machine learning into a recommendation system.

- We develop a new ECPRMML model for recommendation, which obtains superior embeddings of ecological civilization patterns for entities through generation and extraction, extensive experiments demonstrate the effectiveness of our model.

The remainder of this paper is outlined as follows: Section 2 investigates related works. In Section 3, we present the overall framework and implementation process of the ECPRMML method. In Section 4, a series of experiments were conducted to validate the model's performance. In Section 5, we discuss the influence of hyperparameters, multimodal data fusion methods, and data collection methods. Lastly, in Section 6, we summarize our work.

## 2. Related work

In this section, we first present methods tailored for ECPRS and analyse their limitations. We then briefly introduce the applications of Multimodal Machine Learning and LLMs in the field of recommendation systems.

### 2.1 Methods tailored for ECPRS

The first method is KGCN4CEPR(Zeng, Wang, Zhu, Xu, & Zou, 2022)based on a knowledge graph. They utilize a knowledge graph of the regions and employ convolutional operations to extract spatial heterogeneity features of the regions. The convolution process may introduce a significant amount of noise and interfere with recommendation performance. To reduce noise generation, during the convolution operation, some nodes are randomly discarded (H. Wang, Zhao, Xie, Li, & Guo, 2019).

The second method S-GAN (Xu et al., 2022) based on data augmentation, utilizing Generative Adversarial Networks (GANs)(Goodfellow et al., 2014) to learn the distribution of real geographic feature data. This method combines with the third law of geography(Zhu, Lu, Liu, Qin, & Zhou, 2018), "the more similar the geographical environment, the closer the geographical features," to generate numerous fake sample points, alleviating the sparsity of geographic data and enhancing recommendation performance by increasing the quantity of training data. However, it may introduce biases and produce a large number of similar data points, resulting in poor model generalization.

The third method UGPIG (Yu et al., 2023)based on graph pruning and intent network, similar to the first method. Additionally, an intent network is utilized to model the connection signals between regions and ecological civilization patterns, providing some interpretability. Although this method alleviates the sparsity issue of data through the higher-order connectivity of the graph and models collaborative signals between regions and ecological civilization patterns, there is a limitation when embedding ecological civilization patterns. The embedding only uses numerical id for these patterns. Which lead to these commonalities are overlooked during the embedding process.

The above three methods has its own flaws. In the first method, some neighboring nodes may be discarded during convolution, and these discarded neighboring nodes could be crucial elements. The second method, the model has poor generalization capability. The third method, in contrast to regions that use knowledge graphs to get a representation, only the id of the pattern is utilized in the pattern embedding phase to get an embedded representation of the pattern. Using numerical id embedding alone is not sufficient to characterize the ecological civilization pattern.

**2.2 Multimodal recommendations system**

The goal of multimodal machine learning is to correlate and process information from various modalities such as speech, text, and video (Baltrušaitis, Ahuja, & Morency, 2018), utilizing information from different modalities helps provide more comprehensive, enriched, and accurate data representations to support various task. For example, emotion recognition(Y. Zhang, Chen, Shen, & Wang, 2022), speech recognition (Burchi & Timofte, 2023), estimation of redshift for quasars(Hong et al., 2023), and recommendation systems(Wei, Huang, Xia, & Zhang, 2023). A multimodal recommendation system generally consists of three main stages(Q. Liu, Hu, Xiao, Gao, & Zhao, 2023): feature extraction, feature interaction, and recommendation. (1) Feature extraction: Each item has two or more modalities. Modality encoders are used to extract features for each modality, obtaining modality representations in different semantic spaces. For example, models like ViT (Dosovitskiy et al., 2020) can be used to extract image features. (2) Feature interaction: After obtaining modality representations in different semantic spaces, to maximize the effectiveness of features for each modality, features between different modalities are fused in this stage. Common methods include constructing a multimodal knowledge graph(Sun et al., 2020) and utilizing graph neural networks(H. Liu et al., 2020) to aggregate information on the graph. Additionally, simple methods such as direct addition and Multilayer Perceptron are also used. (3) recommendation: After the second step of fusion to obtain representations for users or items, a typical recommendation model can calculate preference scores or probabilities for users towards items. By comparing these, the recommendation results can be obtained.

As a result, multimodal machine learning excels on a variety of tasks including recommendation recommender systems by obtaining richer data representations. Similar to traditional recommendation systems, ECPRS can also benefit from multimodal machine learning, enhancing its performance.

## 2.3 Large Language Models in Recommender Systems

Large language models have become one of the most powerful tools in the field of Natural Language Processing (NLP). Leveraging their strong generalization abilities, they have gained widespread applications in the domain of deep learning. Similarly, large language models have garnered significant attention in the field of Recommendation Systems (RS). Over the recent years, significant progress has been made in utilizing large language models to enhance recommendation systems. The existing work in this field can be categorized into the following three types(Wu, Zheng, et al., 2023): (1) **LLM Embeddings + RS**: By treating large language models as feature extractors, embeddings are obtained by inputting the features of users or items. These embeddings, derived from the large language model, are then fed into the recommendation model to obtain the recommendation results. For example, employing pretraining and fine-tuning of U-Bert(Qiu, Wu, Gao, & Fan, 2021), with a specific emphasis on supplementing user representations in content-rich domains, thus enhancing the performance of the recommendation system. (2) **LLM Tokens + RS**: This approach generates tokens based on the features of users or items. These tokens encapsulate the users' latent preferences for items, providing valuable assistance in the decision-making process of the recommendation system. For instance，Conversational recommender system model UniCRS(Xiaolei Wang, Zhou, Wen, & Zhao, 2022) is designed with three types of tokens: dialogue context, fused knowledge representations, and task-specific soft tokens to provide sufficient semantic information to assist recommendation. (3) **LLM as RS**: This method differs from the two previous decision-assistive approaches. It directly utilizes LLMs to generate appropriate recommendation results based on prompts and Tuning. It has been extensively applied in various domains such as Sequential Recommendation(Sileo, Vossen, & Raymaekers, 2022), Group Recommendation(S. Zhang, Zheng, & Wang, 2022), Explainable RS(Li, Zhang, & Chen, 2023) and Job Recommendation(Wu, Qiu, Zheng, Zhu, & Chen, 2023). Therefore, this approach is also the most commonly used method.

Different from the aforementioned methods, this study introduces the utilization of the powerful comprehension and generation capabilities of LLMs. This allows LLMs to generate multimodal features related to ecological civilization patterns. By extracting these features to enrich the representation of patterns, the performance of the recommendation system is thereby enhanced.

## 3. Method

In this section, we firstly formulate the ecological civilization pattern recommendation task. Following that, we introduce the framework of our model, ECPRMML. Finally, we provide a detailed description of the implementation details of the ECPRMML method.

### 3.1 Problem Formulation

We have several target regions, organized at the district or county level, such as Qixia District, Xifeng District, etc. Additionally, there are several established ecological civilization patterns, such as the Beautiful Countryside pattern, National Park pattern, and so on. Our task is to recommend suitable ecological civilization patterns for these regions based on their characteristics and needs. Let $R$ represent the set of target regions, $P$ represent the set of ecological civilization patterns, and matrix $Y \in \mathbb{R}^{|R| \times |P|}$ represent the interaction matrix. If $y_{r,p} = 1$, it indicates that region r is developing ecological civilization pattern p, meaning there is an interaction. $y_{r,p} = 0$ indicates no interaction. Furthermore, $KG$ is a knowledge graph constructed based on the target regions and their geographical features. $PI$ represents the image data of ecological civilization patterns, and $PT$ represents the text data of ecological civilization patterns.

Our goal is to learn a function $\mathcal{F}$ based on the historical interaction between $R$ and $P$, the provided KG, data $PI$ and $PT$. Through this function, we can calculate the score $\hat{S}_{r,p}$ for each target region r with respect to each ecological civilization pattern p, as follows:

$$\hat{S}_{r,p} = \mathcal{F}(r, p | KG, PI, PT, Y) \#(1)$$

The magnitude of the score represents the degree of preference that $r$ has for $p$. By comparing the scores of one region for all elements in the set $P$, we can determine the most suitable ecological civilization patterns for the development of that region.

### 3.2 Framework Overview

Figure 2 illustrates the overall framework of our method ECPRMML. As shown in Figure 2, the ECPRMML method can be divided into three stages: region feature extraction, pattern feature extraction, and pattern prediction. Firstly, in the region feature extraction stage, we construct a knowledge graph KG based on the regions and their geographical features. KG contains connections such as <region 1, feature, region 2>, where different regions are linked through similar geographical features, indicating a high-order connectivity among regions. By performing aggregation operations on the knowledge graph, we can capture high-order connectivity signals among different regions, resulting in spatial heterogeneity features. Secondly, in the pattern feature extraction stage, we use a large language model like ChatGPT to generate definitions $PT$ for ecological civilization patterns, and we use another one to generate images or manually crawl images $PI$ for ecological civilization patterns. Neural networks are then used to extract image and text features for ecological civilization patterns. Finally, in the pattern prediction stage, once we have obtained the representations for regions and patterns, we can calculate the final recommendation results.

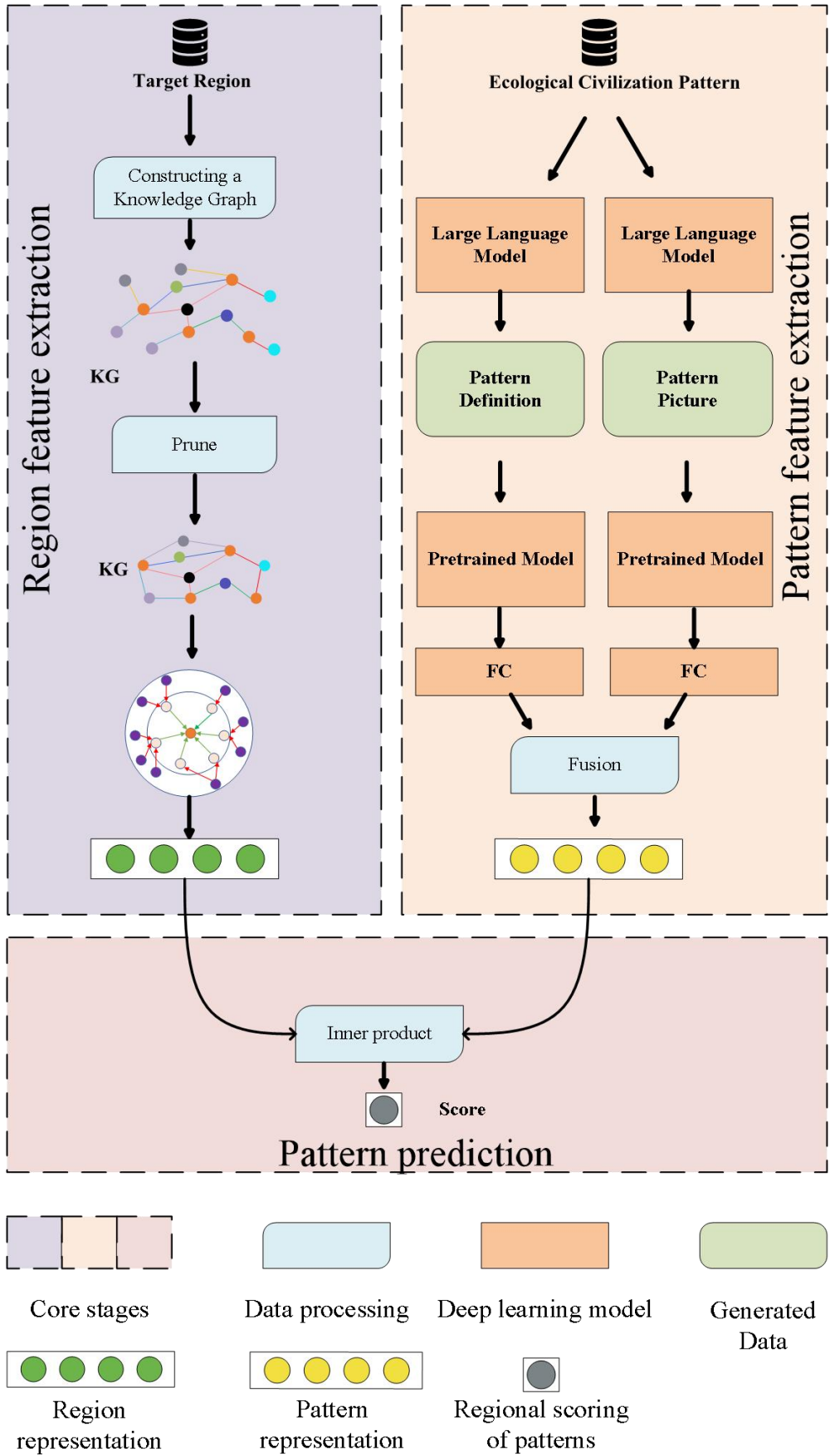

Figure 2 The structure of ECPRMML

### 3.3 Region Feature Extraction

The spatial heterogeneity features of different regions are hidden within the geographical features of these regions. To uncover these features, we have constructed a knowledge graph, which is a complex data structure capable of accommodating and representing spatially relevant information. Part of the constructed knowledge graph is shown in Figure 3. By leveraging the rich semantic information and high-order connectivity signals inherent in the knowledge graph, we can effectively extract these features.

We have designed a total of 29 categories of geographical features from various aspects such as society, culture, infrastructure, resources, economy, and environment to ensure that the content in the knowledge graph is rich enough. At this point, including the regions themselves, there are a total of 30 types of nodes in the knowledge graph. The region nodes are connected to their geographical features through 29 different types of relationships, and ultimately, all nodes form a large semantic network. It should be noted that, although the constructed knowledge graph is of considerable size, there are many isolated feature nodes within the knowledge graph. These nodes are connected only to the region they belong to, resulting in a low graph density. The reason for this is that many features are represented as continuous numerical values that are not shared by many regions. Therefore, after constructing the knowledge graph, a series of pruning operations are performed to increase the graph's density and enhance the connection between nodes.

After pruning the knowledge graph, we search for nodes' N-hop neighbors on the pruned knowledge graph. N is a hyperparameter. Through aggregation operations on the nodes of the knowledge graph, we ultimately obtain vector representations for the target regions, known as the $RegioRepresentation(RR)$. These representations include spatial heterogeneity features. For detailed information about the construction, pruning, aggregation, and other operations on the knowledge graph, please refer to UGPIG.

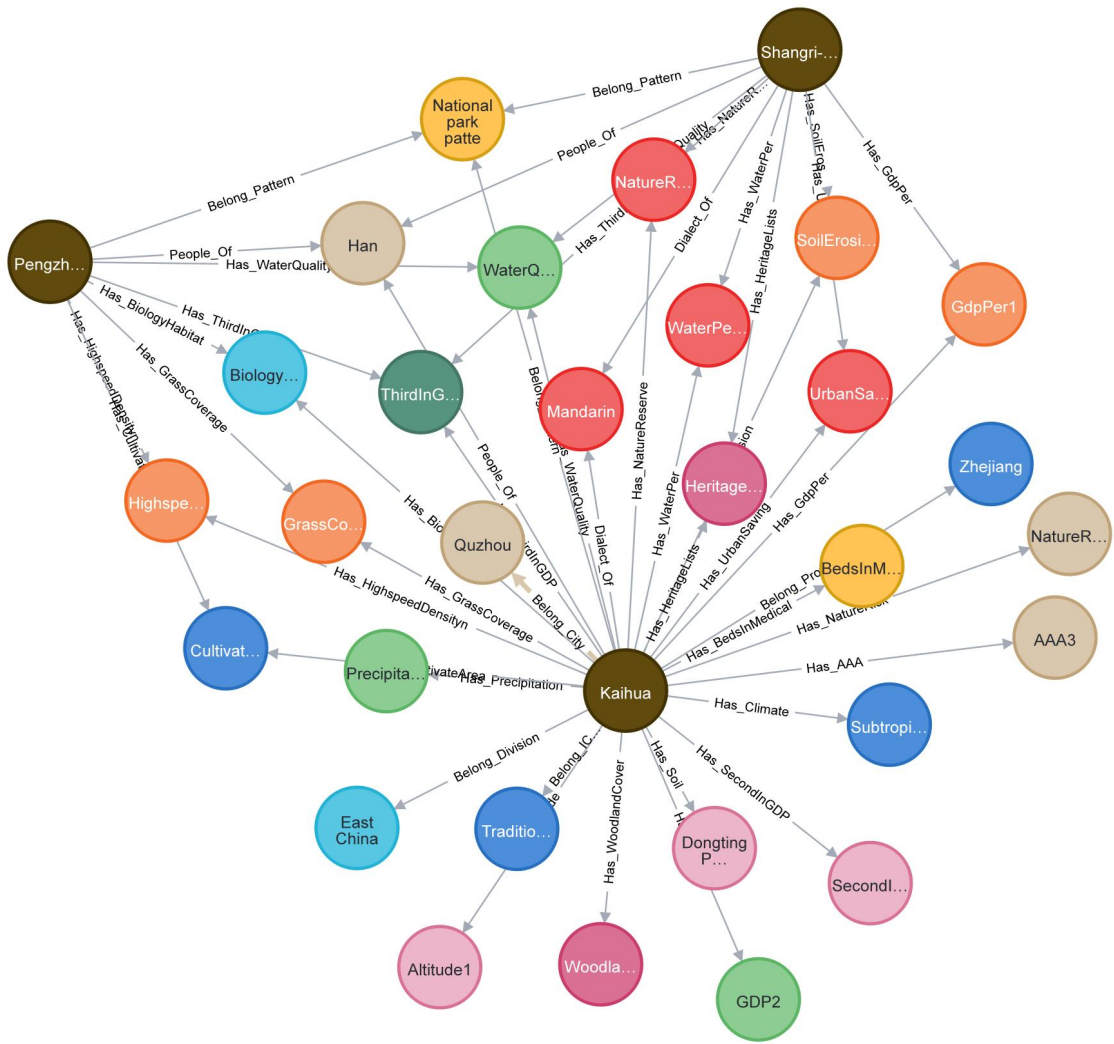

Figure 3 Partial knowledge graph of ECPRMML

## 3.4 Pattern Feature Extraction

In this section, we will separately introduce the process of extracting image features and text features for the patterns.

### 3.4.1 Text Feature Extraction

Text features are important characteristics of ecological civilization patterns. Through text, we can gain a deeper understanding of the principles and implementation methods of ecological civilization patterns. But ecological civilization patterns are a new and abstract concept that does not come with an official or expert-provided definition or textual explanation. In recent years, NLP models have made rapid advancements, demonstrating excellent performance in various text-related tasks, thanks to their strong language understanding and generation capabilities. Inspired by this, in order to obtain definitions for ecological civilization patterns, we use large language model to understand and generate definitions for ecological civilization patterns.

The BERT(Kenton & Toutanova, 2019) model has excellent contextual understanding capabilities, which are very helpful for a deep understanding of the context and semantics of ecological civilization pattern definitions. Therefore, we use a pre-trained Bert-base-Chinese model provided by the Hugging Face community[1] to extract text features from the patterns. Therefore, the extraction process is shown in Figure 4, this process can be divided into three parts:(1) Definition Generation: The ecological civilization pattern data $P$ is input into the large language model

---
[1] https://huggingface.co/

ChatGPT3.5 based on GPT-3(Brown et al., 2020). ChatGPT understands the names of ecological civilization patterns and generates definitions for ecological civilization patterns. (2) Feature Extraction：Once we have obtained all the pattern definitions, they are fed into the Bert-base-Chinese model. Each definition is padded to the maximum input length for BERT. After extraction, each definition is transformed into a 768-dimensional vector called 'pooler_output.' This vector is considered the semantic feature vector for the entire sentence, which is what we desire as the text features. (3) Spatial Mapping: we need to feed the 768-dimensional vector into a fully connected layer with a size of 768×embedding size to obtain the final $Text\ Feature(TF)$. This is done to facilitate the subsequent fusion of features from the two modalities, ensuring that both types of vectors exist in the same vector space.

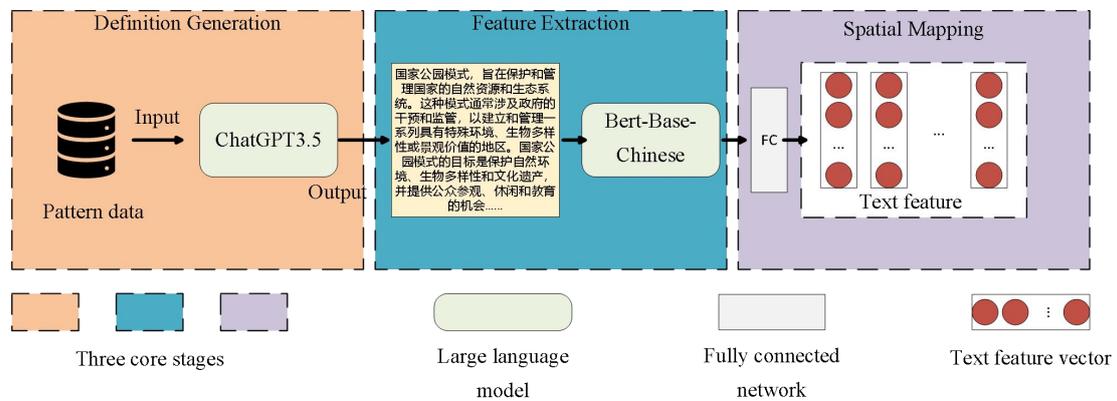

Figure 4 Text feature extraction process

### 3.4.2 Image Feature Extraction

Images can provide the most intuitive reflection of the implementation and impact of ecological civilization patterns. They can convey information about the ecological environment, social culture, economic development, and more in a certain way, helping humans or machines better understand the relationships between them. Therefore, image features are crucial for recommending ecological civilization patterns.

In this work, image sources can be categorized into two main methods: manual crawling and AI generation. The former involves collecting a large number of images related to ecological conservation projects, sustainable development practices, and natural landscapes from the internet using techniques such as web scraping. For each pattern, the most matching image is manually selected from this collection. The latter method utilizes Microsoft's Bing Image Creator service, which generates images based on the names of the patterns. Bing Image Creator is built on OpenAI's LLM DALL·E 2(Ramesh, Dhariwal, Nichol, Chu, & Chen, 2022). By using diverse image data from these sources, we can better support the recommendation and understanding of ecological civilization patterns.

The extraction of image features can be divided into three stages (Figure 5),（1）Image Preprocessing: Initially, the images are resized to 256×256 pixels while maintaining their original aspect ratio. Then, a center crop of 224×224 pixels is taken from the image. Finally, the image is normalized using the mean and standard deviation from ImageNet, with mean= [0.485, 0.456, 0.406] and std= [0.229, 0.224, 0.225]. This standardization is necessary when extracting image features using pre-trained models. (2) Feature Extraction: We will use the pre-trained ResNet-50 model from ImageNet, which has a final fully connected layer of size 2048×1000 used for classification. However, we will modify this layer to be 2048×2048 to extract features without performing a classification task. The preprocessed images from step 1 are fed into the modified ResNet-50(He, Zhang, Ren, & Sun, 2016), resulting in a one-dimensional image feature vector of dimension 2048. (3) Spatial Mapping: Similar to the text features, these image features are passed through a fully connected layer with dimensions 2048×embedding size to obtain the final $Image\ Feature(IF)$.

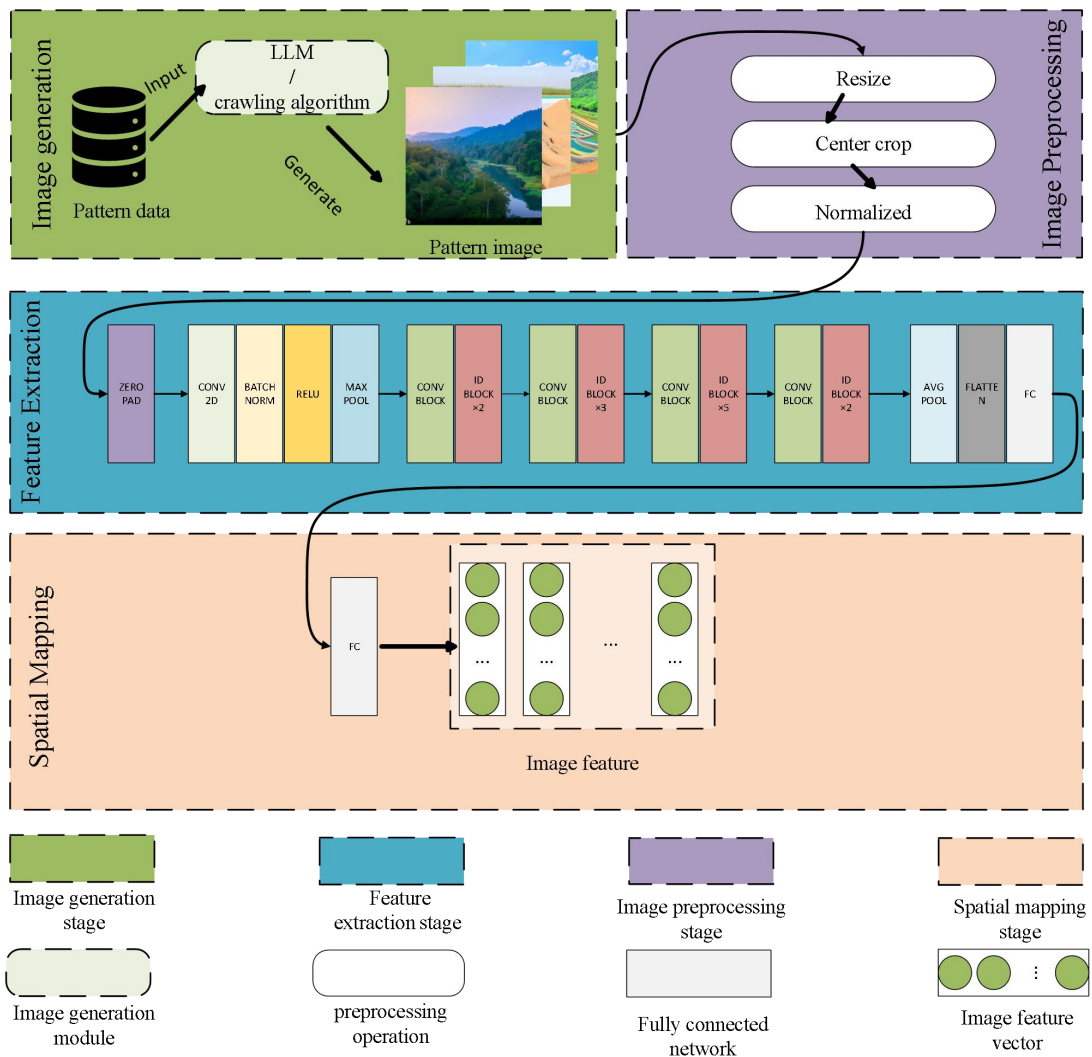

Figure 5 Text feature extraction process

### 3.4.3 Cross-Modal Fusion

The final step of Pattern Feature Extraction is the fusion of features from two modalities to obtain $Pattern\_representation(PR)$ of ecological civilization patterns. Below are several commonly used fusion methods:

**SumFusion** is the simplest method. Assuming both modalities exist in the same vector space, we directly add them together to obtain the final representation of the ecological civilization pattern:

$$PR = IF + TF \#(2)$$

**ConcatFusion** means that during the feature extraction process, without going through the final fully connected layer mentioned in 3.4.1 and 3.4.2, we directly stack the two types of features and then pass them through a fully connected layer:

$$PR = fc(concat(IF, TF)) \#(3)$$

Here $concat$ means concatenating the two features along dimension 1, and $fc$ represents a fully connected layer with a size of 2816×embedding size.

**GatedFusion** (Kiela, Grave, Joulin, & Mikolov, 2018) calculates a gate using one modality, and then uses this gate to modulate the output of the other modality to represent the final fusion result, for example:

$$gate = Sigmoid(IF) \#(4)$$

$$PR = Mul(gate, TF) \#(5)$$

$Sigmoid$ is a common activation function, and $Mul$ represents element-wise multiplication of two vectors. The positions of Image feature and Text feature can be interchanged.

**AttentionFusion** utilizes an attention mechanism to calculate the importance of both modalities and then adds the features of the two modalities according to their importance:

$$PR = \alpha\, IF + \beta\, TF \#(6)$$

$\alpha$ and $\beta$ represent the importance calculated through the attention mechanism, respectively.

We name the methods that use these fusion techniques in ECPRMML as follows: ECPRMML-Sum, ECPRMML-Concatenate, ECPRMML-Gate, and ECPRMML-AttentionSum. The fusion method is a key aspect of the ECPRMML method, and we will evaluate these various methods in our experiments.

### 3.5 Pattern Prediction

In Section 3.3, we obtained the vector representation of regions, denoted as $Region\_representation$, and in Section 3.4, we obtained the vector representation of ecological civilization patterns, denoted as $Pattern\_representation$. After obtaining both of them, the score $\hat{S}_{r,p}$ can be defined as follows:

$$\hat{S}_{r,p} = RR \cdot PR \#(7)$$

Where · represents the inner product symbol, by calculating the score $\hat{S}_{r,p}$ for a specific region and any ecological civilization pattern, and by comparing the scores, we can determine the most suitable ecological civilization pattern for the development of that region.

### 4. Experiments

In this section, we validate the effectiveness of the ECPRMML method on real data. we introduce the dataset used in the experiments, the experimental setup, and the evaluation metrics, then present the baselines and conduct comparative experiment and ablation experiment.

**4.1 Dataset**

This study was conducted in China, with counties/districts serving as the experimental units. Through web text mining, data were collected from 2,596 counties/districts, encompassing 29 geographical features related to culture, society, economy, and more. Additionally, the study collected cases of ecological civilization patterns from these counties, involving a total of 93 distinct ecological civilization pattern instance. Based on the geographical features of 2596 target regions, we constructed a knowledge graph, the scale of which is depicted in Table 1. In section 3.4, the scale of multimodal data for the generated 94 ecological civilization patterns can be found in Table 2. Subsequently, we used the interaction data between target regions and ecological civilization patterns to train our model. For each target region, 80% of the interaction data is allocated to the training set, while the remaining 20% of the data constitutes the test set. Within the training set, 20% of the interaction data is further designated as the validation set. It is important to note that during the training process, the interactions between target regions and the ecological civilization pattern are considered positive samples, while the absence of interaction is considered a negative sample, positive samples are trained along with randomly sampled negative samples to train the model.

Table 1 Statistics of the KG

|  | Regions | Features | Triples | Relationship types |
|---|---|---|---|---|
| number | 2596 | 1669 | 78559 | 29 |

Table 2 Statistics of the multimodal data

|  | Pattern definition | Pattern picture (AI-generated) | Pattern picture (crawled) |
|---|---|---|---|
| number | 12257 (characters) | 94 | 94 |

### 4.2 Evaluation criterion

**Evaluation criterion**: Our goal is to infer which ecological civilization patterns will interact with a given region in the future based on existing interaction data. To assess the performance of our model, we adopt an all-ranking strategy. In other words, we use a portion of interactions to predict whether the remaining interactions will be present in the recommended top-K results. Additionally, due to the sparsity of geospatial data, we set K=5 in this context to facilitate experimental results. We continue to use the same three evaluation metrics as in UGPIG: Precision@K, Recall@K, and F1@K. These are also the three most widely used metrics in recommendation systems. Precision@K represents the proportion of correct results among all the results. In other words, Precision@K assesses how many of the top-K recommended items are relevant. Recall@K represents the proportion of relevant results that are correctly predicted among all the relevant results. F1@K represents the harmonic mean of both Precision@K and Recall@K and serves as a comprehensive measure of the recommendation system's performance. The calculation ways for these three metrics are as follows:

$$Precision@K = \frac{TP@K}{TP@K + FP@K} \#(8)$$

$$Recall@K = \frac{TP@K}{TP@K + FN@K} \#(9)$$

$$F1@K = \frac{2 \times Precision@K \times Recall@K}{Precision@K + Recall@K} \#(10)$$

**Experimental environment:** All experiments were conducted using the PyTorch framework, specifically including Python 3.8, PyTorch 1.12.0, NumPy 1.21.5, and a 12th Gen Intel (R) Core (TM) i5-12500H processor with a clock speed of 3.10 GHz. The GPU used for these experiments is the NVIDIA GeForce RTX 3050 Ti Laptop GPU.

**Parameter Settings**: After fine-tuning on the validation set, we obtained the optimal set of parameters for our ECPRMML method. We set the dimension of the vector embeddings to 64, the number of layers for knowledge graph aggregation to 3, the learning rate to 1e-3, and used manually collected images rather than those generated by large models.

**Baselines**：To validate the performance of the ECPRMML method in the context of ecological civilization pattern recommendation, we conducted multiple sets of experiments on three baseline models: CKE(F. Zhang, Yuan, Lian, Xie, & Ma, 2016), NGCF(Xiang Wang, He, Wang, Feng, & Chua, 2019), UGPIG, and various variants of ECPRMML.

•**CKE** is a method that combines knowledge graph and collaborative filtering. It utilizes the TransR (Lin, Liu, Sun, Liu, & Zhu, 2015) model to extract semantic features from structured knowledge in the knowledge graph. These extracted features from the knowledge graph are then integrated into collaborative filtering to enhance matrix factorization.

•**NGCF** models the high-order connectivity between users and items. It leverages high-order connectivity to associate users and items and uses the idea of Graph Convolutional Networks (GCN)(Kipf & Welling, 2016) to extract useful information at each layer. This effectively incorporates collaborative signals between users and items into the entity embedding process, ultimately enhancing the performance of recommendation systems.

•**UGPIG** utilizes a knowledge graph to capture the spatial heterogeneity features of various target regions. It uses an intent network to extract interaction signals between regions and ecological civilization patterns, combining both to enrich the entity representations of target regions and improve recommendation accuracy.

•**ECPRMML-S** stands for a variant of ECPRMML that solely relies on spatial heterogeneity features between regions to construct the model, without considering the text features and image features of ecological civilization patterns.

•**ECPRMML-I** exclusively utilizes the image features of ecological civilization patterns to construct the model, without taking into account the text features and spatial heterogeneity features of regions.

•**ECPRMML-T** only utilizes the text features of ecological civilization patterns to construct the model, without considering the image features and the spatial heterogeneity features of regions.

•**ECPRMML-SI** incorporates both spatial heterogeneity features between regions and the image features of ecological civilization patterns to construct the model, without considering the text features of ecological civilization patterns.

•**ECPRMML-ST** utilizes both spatial heterogeneity features between regions and the text features of ecological civilization patterns to construct the model, without considering the image features of ecological civilization patterns.

•**ECPRMML-IT** uses both the text features and image features of ecological civilization patterns to construct the model, without considering the spatial heterogeneity features between regions.

**4.3 Results**

In Section 4.3, we set K=5 to evaluate the Top-5 recommendation performance of the aforementioned baselines and the ECPRMML method. The comparative experiment results are shown in Table 3, and it is evident that our method outperforms the others in various evaluation metrics. The results of ablation experiments are presented in Table 4, which demonstrates the significance of each modality of data and the knowledge graph in the recommendation process.

From Table 3, we can see that overall, our ECPRMML method outperforms all three baseline models in all evaluation metrics. Specifically, in terms of $Precision@5$, it achieved a maximum improvement of 1.24%, which is a 0.67% increase compared to the best-performing baseline model. with respect to $Recall@5$, it achieved a maximum improvement of 6.69%, which is a 3.78% increase compared to the best-performing baseline model. As for $F1@5$, it achieved a maximum improvement of 2.11%, which is a 1.16% increase compared to the best-performing baseline model. Analyze the reasons for its improvement，although CKE has introduced heterogeneous information from the knowledge base to enhance the recommendation quality of the recommendation system, it overlooks the rich geographical features of each target region, thereby neglecting spatial heterogeneity. NGCF improves recommendations by using interaction signals between the target region and ecological civilization patterns, but it disregards the characteristics of the target region and ecological civilization patterns. UGPIG, on the other hand, takes into account the spatial heterogeneity features but ignores the characteristics of the ecological civilization patterns themselves. Therefore, we can summarize the reasons for the improvement as follows: (1) The introduction of multimodal data enriches the embedding representation of ecological civilization patterns in the model. The incorporation of both text and image data types allows the model to better explore the nuances of ecological civilization patterns. (2) Utilizing knowledge graphs to extract features of target regions and representing the features of ecological civilization patterns with multimodal data not only results in better embedding representations, but also enhances the description of the relationships between target regions and ecological civilization patterns.

Table 3 Top-5 recommendation result of comparative experiment (Bold font is our method).

|  | $Precision@5$ | $Recall@5$ | $F1@5$ |
| --- | --- | --- | --- |
| CKE | 0.0757(-1.24%) | 0.3004(-6.55%) | 0.1209(-2.11%) |
| NGCF | 0.0765(-1.16%) | 0.2990(-6.69%) | 0.1218(-2.02%) |
| UGPIG | 0.0814(-0.67%) | 0.3281(-3.78%) | 0.1304(-1.16%) |
| **ECPRMML** | **0.0881** | **0.3659** | **0.1420** |

In the ablation experiments, we refer to the features extracted from the knowledge graph as spatial heterogeneity features (S). Additionally, we employ neural networks to extract text features (T) and image features (I). Firstly, as we can see in Table 4, the ECPRMML method, which combines all three types of features, achieves the best performance. Secondly, except for ECPRMML-IT, we can observe that as the number of features increases, the performance also improves, indicating that the inclusion of features leads to an enhancement in the recommendation system's performance. Finally, from the model experiments comparing individual features, it can be seen that the importance of image features is greater than spatial heterogeneity features, and spatial heterogeneity features are greater than text features. We argue that images encompass many aspects such as color, texture, shape, and spatial relationships, making them the most important. The reason for text contributing the least to improvement is twofold: On the one hand, it does not represent the true definition of ecological civilization patterns but is generated by a large language model ChatGPT. On the other hand, textual information is not as rich as images in terms of content.

Table 4 Top-5 recommendation result of ablation experiment (Bold font is our method).

|  | $Precision@5$ | $Recall@5$ | $F1@5$ |
| --- | --- | --- | --- |
| ECPRMML-SI | 0.0814(-0.67%) | 0.3468(-1.91%) | 0.1318(-1.02%) |
| ECPRMML-ST | 0.0745(-1.36%) | 0.3256(-4.03%) | 0.1212(-2.08%) |
| ECPRMML-IT | 0.0707(-1.74%) | 0.2912(-7.47%) | 0.1138(-2.82%) |
| ECPRMML-S | 0.0648(-2.33%) | 0.2646(-10.13%) | 0.1041(-3.79%) |
| ECPRMML-I | 0.0724(-1.57%) | 0.3028(-6.31%) | 0.1168(-2.52%) |
| ECPRMML-T | 0.0374(-5.07%) | 0.1598(-20.6%) | 0.0606(-8.14%) |
| **ECPRMML** | **0.0881** | **0.3659** | **0.1420** |

## 5. Discussion

In section 5, we discuss the impact of various aspects on the model's performance. These aspects included model hyperparameters, modalities fusion methods, and the influence of generated images versus real images on the model's performance.

**5.1 Discussion on Selecting Parameters**

In this section, we used the optimal parameter settings and systematically varied a single parameter while keeping the others fixed. This allowed us to observe how changes in that specific parameter affected the overall performance of the recommendation system, as measured by the comprehensive metric F1@5 metric.

**Impact of embedding dimensions**: As shown in Figure 6(a), we conducted experiments with different embedding dimensions to study their impact on recommendation results. Ultimately, we found that the most suitable embedding dimension was 64, as it effectively represented the information encapsulated by the entities.

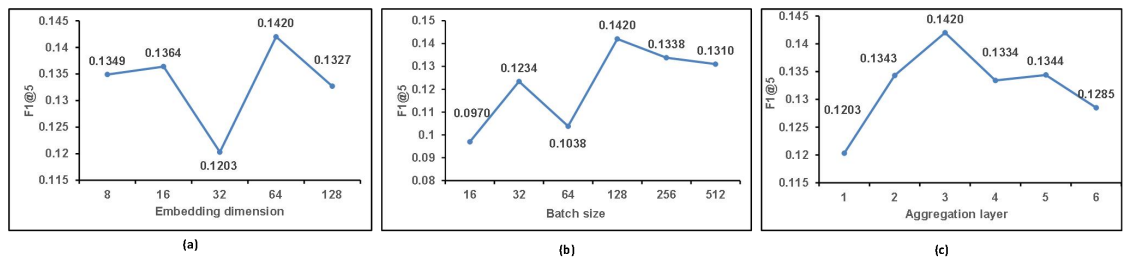

Figure 6 Impact of embedding dimensions

**Impact of batch size**: In our experiments, we set the batch size to be a power of 2, gradually increasing it. As can be seen in Figure 6(b), the recommendation system's performance reached its peak when the batch size was set to 128.

**Impact of number of aggregation layer**：The number of layers for knowledge graph aggregation is a critical parameter, and as shown in Figure 6(c), we can observe that initially, as the number of aggregation layers increases, the recommendation performance also improves. However, after reaching a critical point, the performance starts to decline. This aligns with our initial expectations: as the number of aggregation layers increases, more information is gathered through aggregation. However, at the same time, with more aggregation layers, there are more repeated nodes being used, which introduces more noise, ultimately leading to a decrease in performance.

## 5.2 Discussion on Fusion Method

In this section, we will discuss several cross-modal fusion methods mentioned in section 3.4.3.

**Impact of feature fusion method**：As observed from Figure 7, overall, compared to the Sum-based methods, the two modalities Concatenation methods resulted in the worst performance across the Precision@5, Recall@5, and F1@5 metrics. We argue that the dimensions of features from the two modalities do not match, with the image features having significantly higher dimensions compared to text features. This results in an uneven feature space, and the two modalities' features may contain information of different magnitudes or importance. Simply concatenating them together may lead to one modality's information overshadowing the other modality, thereby reducing the quality of the fused features. On the other hand, the method AttentionSum from the UGPIG achieved the best results. This can be attributed to the use of attention mechanisms to extract features from the two modalities based on their differing importance levels.

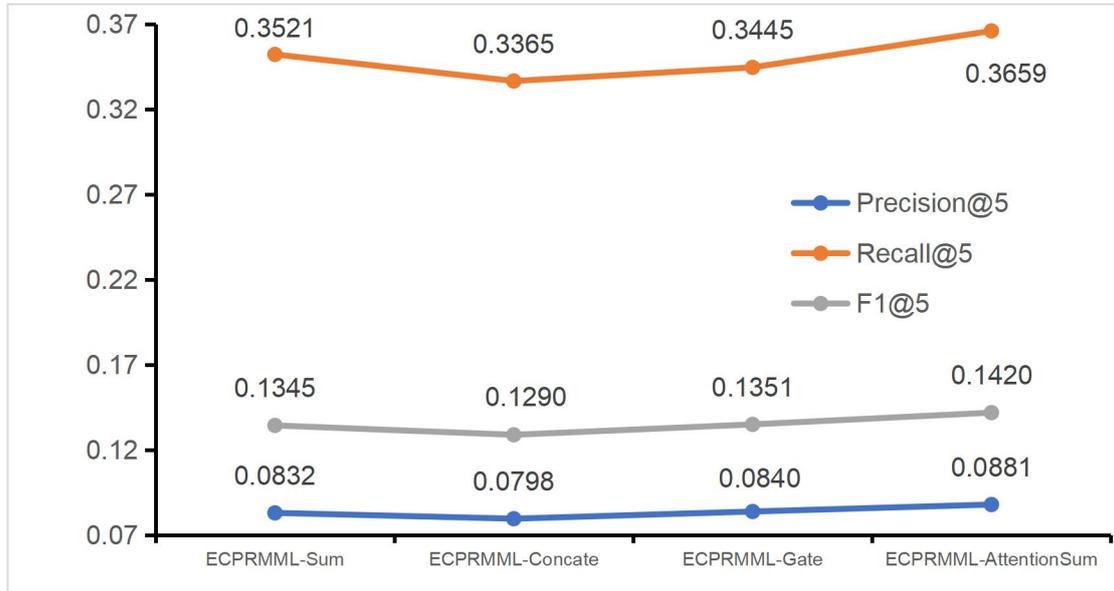

Figure 7 Impact of feature fusion method

## 5.3 Discussion on Generated Images versus Real Images

In this section, we discuss the impact of different types of images on the model. When extracting image features for the ecological civilization pattern, two ways were employed: manual web scraping and automatic image generation using Microsoft's Bing Image Creator service. This discussion focuses on how these different image sources affect the models. Among the studied models involving images, including ECPRMML, ECPRMML-P, ECPRMML-PT, and ECPRMML-SP, experiments were conducted to compare their performance and investigate the influence of the image acquisition method on recommendation performance. The experimental results, as shown in Figure 8, indicate that, except for ECPRMML-I, the overall performance of the other three models is better when using real images. This suggests that in the context of the ecological civilization pattern recommendation scenario, using real image data leads to better performance. Here are the reasons: (1) Abstractness of AI-generated Images: AI-generated images can sometimes be abstract and challenging to identify specific objects or details. (2) Error in AI-generated Images: AI-generated images may contain errors or discrepancies compared to real images, which can impact model performance. These errors can be influenced by both the model itself and the training data. (3) Complexity of the Ecological Civilization Pattern Images: The task of generating and understanding images related to the ecological civilization pattern is inherently complex, posing a challenging task for AI. These factors collectively contribute to the superior performance of models using real images in this particular recommendation context.

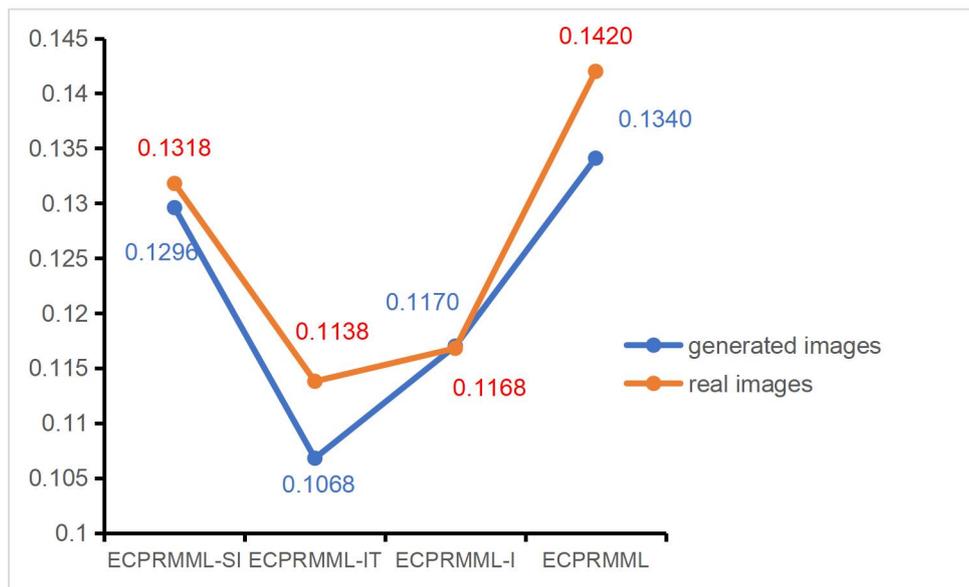

Figure 8 Impact of ways of obtaining images

As shown in Figure 9, these are image examples generated using two different methods, on the left side are generated by the LLM, and on the right side are crawled by a web crawler. Such as waterfowl and aquatic products pattern (Figure 9 (a1)) and (Figure 9 (a2)), on the surface, AI-generated images appear to be brighter, more vibrant, and richer in color. However, upon closer examination, it becomes apparent that some of the ducks in the generated images differ significantly from their actual appearance. Additionally, AI-generated images may contain disruptive elements such as houses and forests, which can introduce disturbances to the model. Hence, this results in the lower performance of the image recommendation system when using images generated by the LLM.

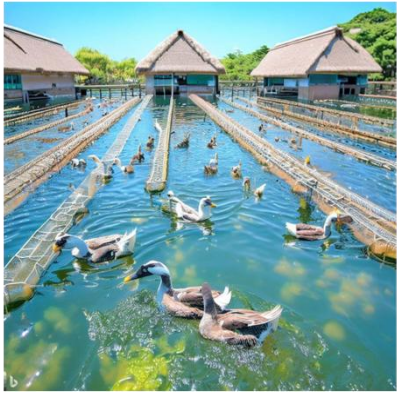 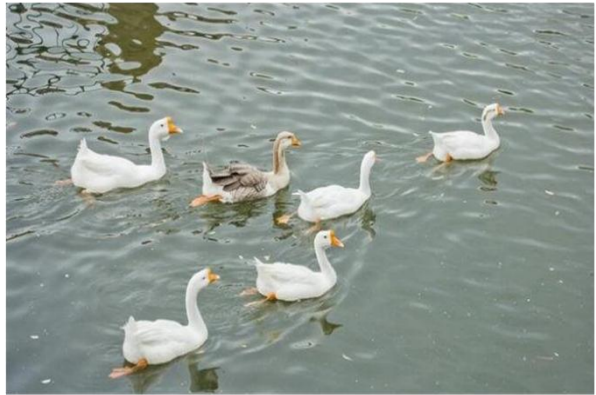

(a1)waterfowl and aquatic products pattern    (a2)waterfowl and aquatic products pattern

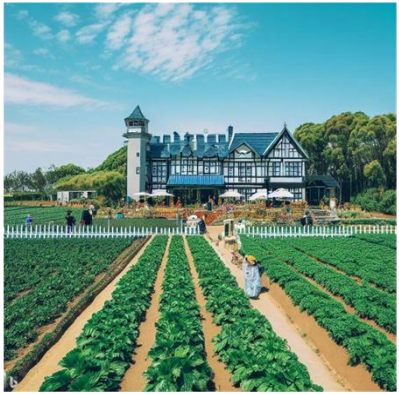 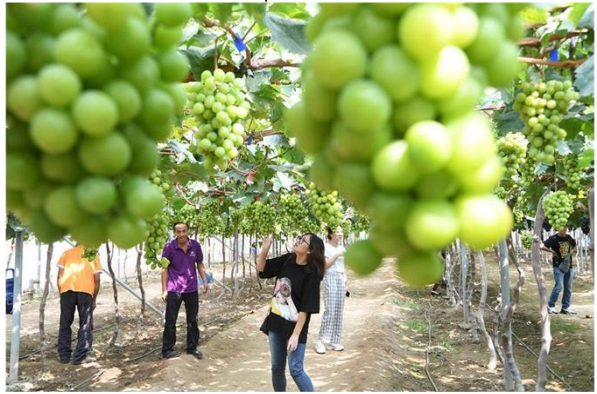

(b1)Estate picking pattern    (b2)Estate picking pattern

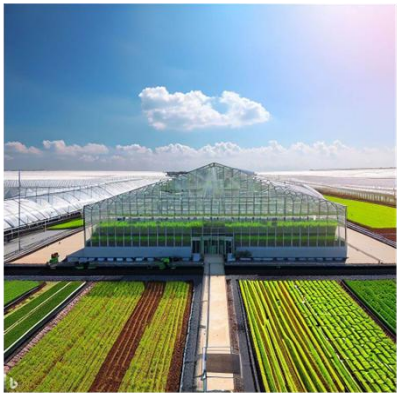 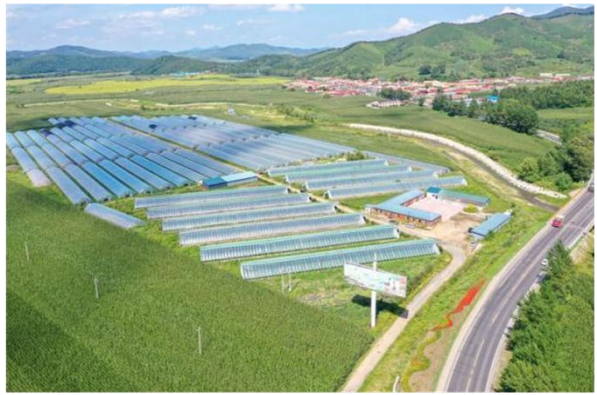

(c1)Facility Agriculture Pattern    (c2)Facility Agriculture Pattern

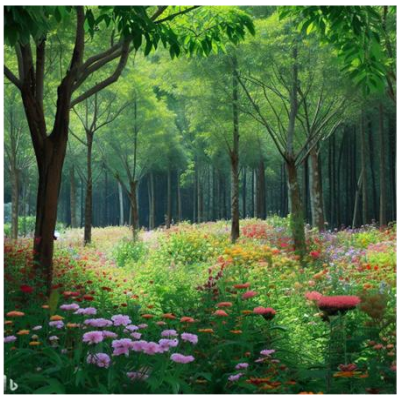 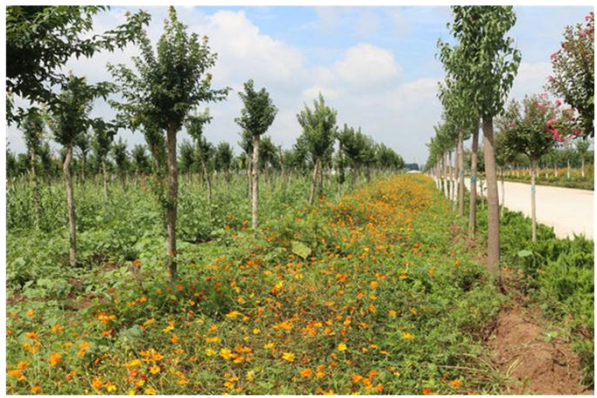

(d1)Flower-Silvicultural Pattern    (d2)Flower-Silvicultural Pattern

Figure 9 Samples of images generated in both ways

## 6. Conclusion

In this paper, we propose a novel model called ECPRMML, which innovatively integrates large language models, knowledge graphs, and multimodal machine learning into a recommendation system. This approach utilizes the knowledge graph to extract region representations containing spatial heterogeneity features. It leverages the generated textual and image features of ecological civilization patterns to compute recommendation scores, providing recommendations for the development of regional ecological civilization patterns. Additionally, through experiments on real datasets, we validate the performance of our proposed method and discuss the impact of information from different modalities on the model's performance.

This work relies on the quality of images and text generated by large language models, as well as the fusion of multimodal features. In future work, we can further improve our ECPRMML model in the following aspects. Firstly, we can design efficient methods for multimodal feature fusion, leveraging graph knowledge and large language models to obtain more expressive feature representations, thereby fully exploring the information from different modalities. Secondly, enhancing the model's interpretability to enable users to understand the basis and reasons behind the recommendations. Additionally, we can introduce generative adversarial networks to improve the quality of generated samples by the model.

# References

Baltrušaitis, T., Ahuja, C., & Morency, L.-P. (2018). Multimodal machine learning: A survey and taxonomy. *IEEE transactions on pattern analysis, 41*(2), 423-443.

Brown, T., Mann, B., Ryder, N., Subbiah, M., Kaplan, J. D., Dhariwal, P., . . . Askell, A. (2020). Language models are few-shot learners. *Advances in neural information processing systems, 33*, 1877-1901.

Burchi, M., & Timofte, R. (2023). *Audio-visual efficient conformer for robust speech recognition.* Paper presented at the Proceedings of the IEEE/CVF Winter Conference on Applications of Computer Vision.

Dosovitskiy, A., Beyer, L., Kolesnikov, A., Weissenborn, D., Zhai, X., Unterthiner, T., . . . Gelly, S. (2020). *An Image is Worth 16x16 Words: Transformers for Image Recognition at Scale.* Paper presented at the International Conference on Learning Representations.

Goodfellow, I., Pouget-Abadie, J., Mirza, M., Xu, B., Warde-Farley, D., Ozair, S., . . . Bengio, Y. (2014). Generative adversarial nets. *Advances in neural information processing systems, 27.*

He, K., Zhang, X., Ren, S., & Sun, J. (2016). *Deep residual learning for image recognition.* Paper presented at the Proceedings of the IEEE conference on computer vision and pattern recognition.

Hong, S., Zou, Z., Luo, A.-L., Kong, X., Yang, W., & Chen, Y. (2023). PhotoRedshift-MML: A multimodal machine learning method for estimating photometric redshifts of quasars. *Monthly Notices of the Royal Astronomical Society, 518*(4), 5049-5058.

Kenton, J. D. M.-W. C., & Toutanova, L. K. (2019). *BERT: Pre-training of Deep Bidirectional Transformers for Language Understanding.* Paper presented at the Proceedings of NAACL-HLT.

Kiela, D., Grave, E., Joulin, A., & Mikolov, T. (2018). *Efficient large-scale multi-modal classification.* Paper presented at the Proceedings of the AAAI conference on artificial intelligence.

Kipf, T. N., & Welling, M. J. a. p. a. (2016). Semi-supervised classification with graph convolutional networks.

Li, L., Zhang, Y., & Chen, L. (2023). Personalized prompt learning for explainable recommendation. *ACM Transactions on Information Systems, 41*(4), 1-26.

Lin, Y., Liu, Z., Sun, M., Liu, Y., & Zhu, X. (2015). *Learning entity and relation embeddings for knowledge graph completion.* Paper presented at the Proceedings of the AAAI conference on artificial intelligence.

Liu, H., Han, J., Fu, Y., Zhou, J., Lu, X., & Xiong, H. (2020). Multi-modal transportation recommendation with unified route representation learning. *Proceedings of the VLDB Endowment, 14*(3), 342-350.

Liu, Q., Hu, J., Xiao, Y., Gao, J., & Zhao, X. (2023). Multimodal Recommender Systems: A Survey. *arXiv preprint arXiv:.03883.*

Qiu, Z., Wu, X., Gao, J., & Fan, W. (2021). *U-BERT: Pre-training user representations for improved recommendation.* Paper presented at the Proceedings of the AAAI Conference on Artificial Intelligence.

Ramesh, A., Dhariwal, P., Nichol, A., Chu, C., & Chen, M. (2022). Hierarchical text-conditional image generation with clip latents. *arXiv preprint arXiv:.06125, 1*(2), 3.

Sileo, D., Vossen, W., & Raymaekers, R. (2022). *Zero-shot recommendation as language modeling.* Paper presented at the European Conference on Information Retrieval.

Sun, R., Cao, X., Zhao, Y., Wan, J., Zhou, K., Zhang, F., . . . Zheng, K. (2020). *Multi-modal knowledge graphs for recommender systems.* Paper presented at the Proceedings of the 29th ACM international conference on information & knowledge management.

Wang, H., Zhao, M., Xie, X., Li, W., & Guo, M. (2019). *Knowledge graph convolutional networks for recommender systems.* Paper presented at the The world wide web conference.

Wang, X., He, X., Cao, Y., Liu, M., & Chua, T.-S. (2019). *Kgat: Knowledge graph attention network for recommendation.* Paper presented at the Proceedings of the 25th ACM SIGKDD international conference on knowledge discovery & data mining.

Wang, X., He, X., Wang, M., Feng, F., & Chua, T.-S. (2019). *Neural graph collaborative filtering.* Paper presented at the Proceedings of the 42nd international ACM SIGIR conference on Research and development in Information Retrieval.

Wang, X., Zhou, K., Wen, J.-R., & Zhao, W. X. (2022). *Towards unified conversational recommender systems via knowledge-enhanced prompt learning.* Paper presented at the Proceedings of the 28th ACM SIGKDD Conference on Knowledge Discovery and Data Mining.

Wei, W., Huang, C., Xia, L., & Zhang, C. (2023). *Multi-Modal Self-Supervised Learning for Recommendation.* Paper presented at the Proceedings of the ACM Web Conference 2023.

Wu, L., Qiu, Z., Zheng, Z., Zhu, H., & Chen, E. (2023). Exploring large language model for graph data understanding in online job recommendations. *arXiv preprint arXiv:.05722*.

Wu, L., Zheng, Z., Qiu, Z., Wang, H., Gu, H., Shen, T., . . . Liu, Q. (2023). A Survey on Large Language Models for Recommendation. *arXiv preprint arXiv:.19860*.

Xu, M., Wang, S., Song, C., Zhu, A., Zhu, Y., & Zou, Z. (2022). The Recommendation of the Rural Ecological Civilization Pattern Based on Geographic Data Argumentation. *Applied Sciences, 12*(16), 8024.

Yu, Z., Wang, S., Zhu, Y., Yuan, W., Dai, X., & Zou, Z. (2023). Unveiling Optimal SDG Pathways: An Innovative Approach Leveraging Graph Pruning and Intent Graph for Effective Recommendations. *arXiv preprint arXiv:2309.11741*.

Zeng, X., Wang, S., Zhu, Y., Xu, M., & Zou, Z. (2022). A Knowledge Graph Convolutional Networks Method for Countryside Ecological Patterns Recommendation by Mining Geographical Features. *ISPRS International Journal of Geo-Information, 11*(12), 625.

Zhang, F., Yuan, N. J., Lian, D., Xie, X., & Ma, W.-Y. (2016). *Collaborative knowledge base embedding for recommender systems.* Paper presented at the Proceedings of the 22nd ACM SIGKDD international conference on knowledge discovery and data mining.

Zhang, S., Zheng, N., & Wang, D. (2022). *GBERT: Pre-training User representations for Ephemeral Group Recommendation.* Paper presented at the Proceedings of the 31st ACM International Conference on Information & Knowledge Management.

Zhang, Y., Chen, M., Shen, J., & Wang, C. (2022). *Tailor versatile multi-modal learning for multi-label emotion recognition.* Paper presented at the Proceedings of the AAAI Conference on Artificial Intelligence.

Zhu, A. X., Lu, G., Liu, J., Qin, C. Z., & Zhou, C. (2018). Spatial prediction based on Third Law of Geography. *Annals of GIS, 24*(4), 225-240.